\documentclass[journal]{IEEEtran}

\usepackage{amsmath, amssymb,booktabs}
\usepackage{tikz,pgfplots,pgfplotstable}
\usepackage{color, colortbl}
\definecolor{Gray}{gray}{0.9}
\pgfplotsset{compat=newest}
\usetikzlibrary{shapes,arrows,positioning,automata, patterns,fit,shadows,decorations.shapes, decorations.markings,decorations.pathreplacing}
\tikzstyle{block} = [rectangle, draw, minimum width=4.5em, text centered, minimum height=2em, inner sep=0.75em, align=center, rounded corners]
\tikzstyle{cross}=[path picture={\draw[black](path picture bounding box.south east) -- (path picture bounding box.north west) (path picture bounding box.south west) -- (path picture bounding box.north east);}]
\usepackage[nospace, compress]{cite}
\title{iSEGAN: Improved Speech Enhancement Generative Adversarial Networks}

\author{Deepak~Baby% <-this % stops a space
\thanks{email: deepak.baby@idiap.ch} \\
Idiap Research Institute, Martigny, Switzerland}% <-this % stops a space
\begin{document}
\maketitle

\begin{abstract}
Popular neural network-based speech enhancement systems operate on the magnitude spectrogram and ignore the phase mismatch between the noisy and clean speech signals. Conditional generative adversarial networks (cGANs) show promise in addressing the phase mismatch problem by directly mapping the raw noisy speech waveform to the underlying clean speech signal. However, stabilizing and training cGAN systems is difficult and they still fall short of the performance achieved by the spectral enhancement approaches. This paper investigates whether different normalization strategies and one-sided label smoothing can further stabilize the cGAN-based speech enhancement model. In addition, we propose incorporating a Gammatone-based auditory filtering layer and a trainable pre-emphasis layer to further improve the performance of the cGAN framework. Simulation results show that the proposed approaches improve the speech enhancement performance of cGAN systems in addition to yielding improved stability and reduced computational effort.
\end{abstract}


\section{Introduction} \label{sec:intro}
%\IEEEPARstart{S}{peech} enhancement systems aim to improve the quality and intelligibility of acquired speech signals by removing the artefacts introduced from background noise or other interferences such as room reverberation. Since the performance of several applications such as automatic speech recognition, mobile communication and hearing aids depends on the quality of the captured audio signal, speech enhancement has been an actively researched topic and several methods have been proposed over the past several decades \cite{loizou_book2013,cohen_book2008}.

%Classical speech enhancement algorithms include spectral subtraction \cite{boll1979} and Kalman filtering \cite{grancharov2006}. However, such statistical model-based approaches make assumptions on the properties of speech and noise that are often invalid on realistic recordings. To overcome this shortcoming, several supervised approaches that learn  the speech and noise statistics from a training dataset have been adopted \cite{dbaby_aslp2015, sreenivas1996}. Recent advances in deep neural network (DNN)-based learning architectures are shown to outperform most of the conventional speech enhancement approaches \cite{wang_aslp2014,yong_aslp2015,xugang_is2013,sun_hscma2017,erdogan_book2017}, thanks to their nonlinear structure with multiple hidden layers which enables them to model the complex degradations in the captured speech signal.

\IEEEPARstart{S}{peech} enhancement systems aim to improve the quality and intelligibility of acquired speech signals by removing artefacts caused by background noise or other interferences such as room reverberation. Recently, deep neural network (DNN)-based approaches gained much success in speech enhancement due to their powerful modeling capabilities \cite{wang_aslp2014, yong_aslp2015, xugang_is2013, sun_hscma2017, erdogan_book2017}. 

DNN-based systems are typically trained to estimate a time-frequency (T-F) mask in the range $[0,~1]$, which provides the relative amplitudes of the underlying clean speech and noise signals at every T-F point \cite{wang_aslp2014, dbaby_is2018}. However, these masks modify only the magnitude spectra of the input signal and ignore the phase mismatch between the noisy and clean speech signals \cite{wang_aslp2014,narayanan_irm2013}. Since speech quality can be significantly improved when the clean phase spectrum is known \cite{paliwal_speechcomm2011}, it is worthwhile exploring speech enhancement techniques which preserve phase information. To remedy this phase mismatch problem, this paper investigates the use of generative neural networks which can directly map the raw noisy speech waveform to the underlying clean speech waveform.

In particular, we use generative adversarial networks (GAN) \cite{gan_nips2014} which consists of a generative model or \emph{generator network} ($G$) and a \emph{discriminator network} ($D$) that play a min-max game between each other. $D$ is trained to distinguish the samples generated by $G$ from the real data. $G$, on the other hand, is trained to fool $D$ into accepting its outputs being real. It was demonstrated that GANs can produce realistic image samples \cite{gan_nips2014}.  However, there is no control on the data being generated in such an unconditional generative model. For speech enhancement tasks, we have to control the generated data based on the input noisy data. In order to address this, conditional GANs (cGANs) \cite{cgan} provide an alternative framework in which the model is conditioned to control the data generation process based on input data.

cGANs have recently been shown to yield promising noise suppression performance \cite{santi_segan_is2017,michelsanti_is2017,donahue_ganasr2017,wang_gan2018}. The technique presented in \cite{michelsanti_is2017} is based on the pix2pix architecture \cite{pix2pix} where the cGAN is trained to generate the spectrogram of clean speech given the noisy speech spectrogram and this technique otherwise ignores the phase mismatch problem. The speech enhancement GAN (SEGAN) system proposed in \cite{santi_segan_is2017} is a 1D adaptation of the pix2pix architecture that operates on the raw waveform. However, the performance of cGAN-based models is still worse than conventional magnitude spectral enhancement approaches. In addition, training GANs is complex as it requires finding a Nash equilibrium of a non-convex game between $G$ and $D$ \cite{gan_nips2014,salimans_nips2016}, and these prior works do not provide much insights on how to achieve this equilibrium.

This paper makes use of the SEGAN model \cite{santi_segan_is2017} as the baseline system and systematically investigates approaches to further stabilize cGAN-based speech enhancement training and improve its performance. The SEGAN setting made use of virtual batch normalization (VBN) \cite{salimans_nips2016} in $D$ to stabilize the training which is computationally expensive in memory and training time. Instead, we propose using instance normalization \cite{instancenorm} which requires fewer computational resources. In addition, we investigate the use of one-sided label smoothing \cite{salimans_nips2016} to further stabilize the GAN training. Lastly, since the ultimate goal of our system is to improve speech intelligibility, we also propose using a trainable auditory filter-bank layer based on Gammatone filter-banks that approximates the cochlear processing in both $G$ and $D$. 

The contributions of this work are threefold. We Present a cGAN-based speech enhancement framework for further research and development, which 1) introduces instance normalization and one-sided label smoothing for training cGAN-based speech enhancement systems, 2) incorporates trainable auditory filtering and pre-emphasis layers to further improve the enhancement quality, and 3) provides an overview of various stabilization methods involved in GAN training.
\section{Speech Enhancement cGAN framework}\label{sec:gan_framework}
\begin{figure}
\input{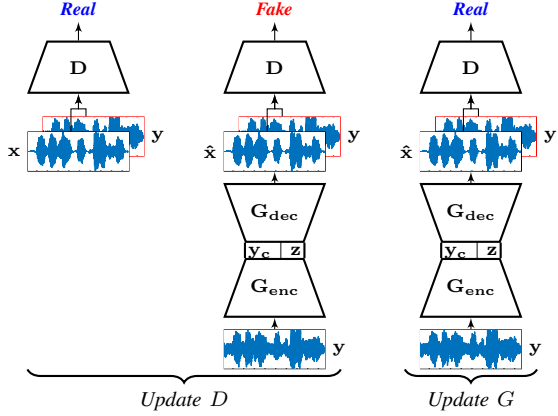}
\caption{Training a cGAN-based speech enhancement system. The updates for $D$ and $G$ are alternated over several epochs. $\mathbf{y}$, $\mathbf{x}$ and $\mathbf{\hat{x}}$ are the noisy speech, clean speech and the clean speech estimate generated by $G$, respectively. $\mathbf{y_c}$ is the encoder output of noisy speech and $\mathbf{z}$ are samples from the prior distribution $\mathcal{Z}$. Due to the adversarial training, $G$ updates its parameters such that it generates samples that are closer to the clean speech manifold.  \label{fig:gan_training}}
\end{figure}

The goal of a speech enhancement system is to estimate the clean speech signal $\mathbf{x}$ from the noisy mixture $\mathbf{y} = \mathbf{x} + \mathbf{w}$, where $\mathbf{w}$ is the added background noise.

In the generic GAN model, $G$ acts as a generative model that learns to map samples $\mathbf{z}$ from some prior distribution $\mathcal{Z}$ to samples $\mathbf{x}$ that belong to a data distribution of interest $\mathcal{X}$ (i.e., the distribution of the clean speech samples, in our case). $D$ is a binary classifier that is trained to classify samples from the true data distribution as real and the generated samples from $G$ as fake. Since $G$ is trained to fool $D$ so that $D$ classifies $G$'s output as real, $G$ will in turn learn to generate samples that are closer to the real data manifold. With cGANs, we direct this data generation process based on the input noisy speech $\mathbf{y}$ such that $G$ generates an estimate that is closer to the underlying clean speech signal $\mathbf{x}$ (denoted as $\mathbf{\hat{x}} \triangleq G(\mathbf{y},\mathbf{z})$).

The training phases of a cGAN-based speech enhancement system are depicted in Fig.  \ref{fig:gan_training}. Notice that $D$ is conditioned using the noisy speech signal $\mathbf{y}$ and $G$ makes use of an encoder-decoder structure. The encoder ($G_{\text{enc}}$) projects the input noisy signal into a condensed representation $\mathbf{y_c} = G_{\text{enc}}(\mathbf{y})$, which is concatenated with the latent samples $\mathbf{z}$. The decoder ($G_{\text{dec}}$) then reconstructs the signal such that its output $\mathbf{\hat{x}} = G_{\text{dec}}(\mathbf{y_c}, \mathbf{z})$ fools $D$ into classifying it as real. As can be seen from Fig. \ref{fig:gan_training}, training a cGAN-based speech enhancement setting is comprised of repeating the following three updates for every mini-batch over several epochs (encoding real as $1$ and fake as $0$):
\begin{enumerate}
\item Update $D$ such that $\mathbf{x}$ and $\mathbf{y}$ pairs are classified as real, i.e., $D(\mathbf{x}, \mathbf{y}) \rightarrow 1$
\item Update $D$ such that the generated samples $\mathbf{\hat{x}}$ and $\mathbf{y}$ pairs are classified as fake, i.e., $D(\mathbf{\hat{x}}, \mathbf{y}) \rightarrow 0$ 
\item Freeze $D$ and update $G$ such that $D$ classifies $\mathbf{\hat{x}}$ and $\mathbf{y}$ pairs as real, i.e., $D(\mathbf{\hat{x}}, \mathbf{y}) \rightarrow 1$
\end{enumerate}

For updating the $G$ and $D$-networks, we use least-squares GAN (LSGAN) \cite{lsgan} which substitutes the conventional cross-entropy loss of the binary classifier $D$ by least-squares. It has been shown that LSGANs further stabilize the GAN training and improve the quality of the generated samples in $G$. In addition, several prior works \cite{santi_segan_is2017,michelsanti_is2017,pix2pix} use an additional loss term that minimizes the L1 distance between the generated samples $\mathbf{\hat{x}}$ and the clean examples $\mathbf{x}$. This L1 term is controlled by a new hyper-parameter $\lambda$. Thus, the loss functions used for updating $D$ and $G$ are,
\begin{align*}
\min_{D} \mathcal{L}(D) &= \dfrac{1}{2}~ \mathbb{E}_{\mathbf{x}, \mathbf{y} \sim p_{\text{data}}(\mathbf{x}, \mathbf{y})} \left[  D(\mathbf{x}, \mathbf{y}) - 1 \right]^{2}  \\
&~~~+ \dfrac{1}{2}~\mathbb{E}_{\mathbf{z}\sim \mathcal{Z},~\mathbf{y} \sim p_{\text{data}}(\mathbf{y})} \left[  D(\mathbf{\hat{x}} , \mathbf{y}) \right]^{2} \\
\min_{G} \mathcal{L}(G) &= \mathbb{E}_{\mathbf{z}\sim \mathcal{Z},~\mathbf{y} \sim p_{\text{data}}(\mathbf{y})} \left[  D(\mathbf{\hat{x}} , \mathbf{y}) -1 \right]^{2} + \lambda \Vert \mathbf{\hat{x}} - \mathbf{x} \Vert_{1}.
\end{align*}

\begin{figure}[t!]
\input{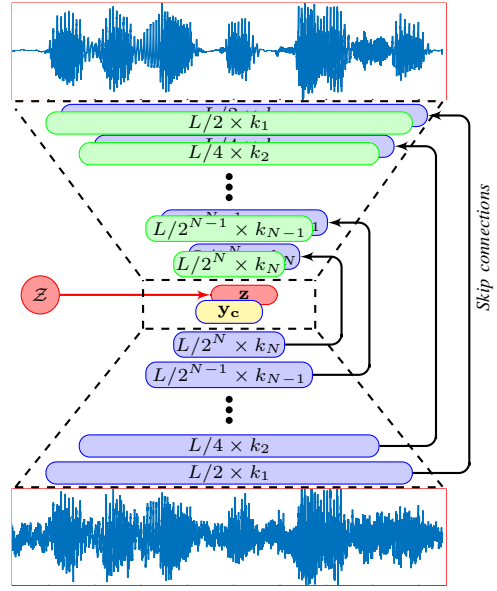}
\caption{Generator architecture: Encoder-decoder structure featuring U-shaped skip-connections employed for speech enhancement. Normalization and activation layers are omitted. The arrows denote the connection between the layers. The output shapes of each layer are also provided, where $L$ and $k_n$ are the length of the input signal and the number of feature-maps at the $n^\text{th}$ layer, respectively. $\mathbf{y_c}$ denotes the encoder output corresponding to the input noisy speech signal and $\mathbf{z}$ are the samples from the prior distribution $\mathcal{Z}$.}
\label{fig:generator}
\end{figure}

\subsection{$G$-network}
In the cGAN-based speech enhancement framework, $G$ performs the enhancement. The $G$ architecture employed is depicted in Fig. \ref{fig:generator}. Similar to prior works \cite{santi_segan_is2017,michelsanti_is2017}, $G$ is designed to be fully convolutional which enforces the network to focus on temporally-close correlations in the input signal. $G_{\text{enc}}$ projects and compresses the input noisy signal through several strided convolutional layers followed by a parametric rectified linear unit (PReLU) non-linearity \cite{prelu}. Strided convolution (stride $>1$) is preferred over other pooling approaches as they provide a more stable GAN training \cite{dcgan}. $G_{\text{dec}}$ uses an inverted version of $G_{\text{enc}}$ by means of fractional-strided deconvolutions, followed by PReLUs. 

Also notice that $G$ uses U-shaped skip-connections that bypass the intermediate compression stages (Fig. \ref{fig:generator}). These skip connections directly pass the fine-grained information such as phase and alignment to the decoder. They also provide a better training behavior as the gradients can flow deeper through the whole network \cite{he_cvpr2016}.

\subsection{$D$-network}
$D$-network makes use of the same structure as $G_{\text{enc}}$, but with a few differences: 1) it has two input channels (one for $\mathbf{x}$ or $\mathbf{\hat{x}}$; and one for $\mathbf{y}$), 2) it uses a normalization layer before the non-linearity, 3) it uses LeakyReLU non-linearity instead of PReLU, and 4) there is an additional convolutional layer with one filter of width $1$ ($1\times 1$ convolution) and its output is fed to a fully-connected layer to perform the binary classification.  

\subsection{Proposed cGAN variants}
\subsubsection{Instance normalization}
An instance normalization layer \cite{instancenorm} applies mean-variance normalization on every channel and input sample. It was successfully used for image stylization \cite{instancenorm} and dehazing \cite{instancenorm_dehaze} and requires less computational complexity than VBN. Motivated by this, we propose to use instance normalization in $D$ for training the cGAN model. To our knowledge, instance normalization has not yet been investigated for cGAN-based speech enhancement. 

\subsubsection{One-sided label smoothing}
One of the critical scenarios which results in unstable GAN training is when $D$ becomes too confident on the real examples, such that $G$ no longer can fool it. One simple trick to remedy this is to encourage $D$ to estimate soft probabilities on real samples, for e.g., $D(\mathbf{x}, \mathbf{y}) \rightarrow 0.9$ instead of $1$. This solution avoids overpowering of $D$ over $G$ and could stabilize GAN training. This approach is called one-sided label smoothing \cite{salimans_nips2016} since only the confidence on real samples is modified.

\subsubsection{Auditory filter-bank layer}
The ultimate goal of speech enhancement systems is to improve speech intelligibility. However, existing cGAN-based systems \cite{santi_segan_is2017,michelsanti_is2017} which are simple adaptations of the pix2pix architecture do not take this goal into account and provide full freedom to $G$ and $D$. In this work, we replace the first layer of both $G$ and $D$ with an auditory filter-bank layer that mimics human auditory processing. Similar ideas have been used for speech recognition applications \cite{hoshen_icassp2015, sainath_is2015}, but have not yet been used in this context. We make use of a Gammatone-based model for the cochlear filtering and use it to initialize the input layers.

\subsubsection{Pre-emphasis layer}
In most speech processing applications, it is beneficial to boost the high-frequency signal content by means of a pre-emphasis filter. This is typically implemented as a first oder high-pass filter $\tilde{y}[n] = y[n] - \alpha y[n-1]$, with $0.9 \leq \alpha < 1$. Instead of using a fixed $\alpha$, we propose to optimize it by implementing the pre-emphasis filter in $G$ as a trainable convolutional layer of filter-length $2$ and stride $1$. The layer is initialized with weights $[-0.95,~1]$ and is trained together with the cGAN network.

\subsubsection{Removing the latent vector $\mathbf{z}$}
It is observed in \cite{pix2pix} that adding the latent vector $\mathbf{z}$ for image processing applications is sometimes not effective as the generator simply learns to ignore it. Some prior works on cGAN-based speech enhancement \cite{donahue_ganasr2017,wang_gan2018} therefore omitted $\mathbf{z}$ such that cGAN generates deterministic outputs. However, it is to date unclear whether $\mathbf{z}$ is helpful for speech enhancement applications. To investigate this, a comparison of all the proposed algorithms with and without $\mathbf{z}$ is included.
\section{Evaluation Setup \label{sec:eval_setup}}

\subsection{Database}
The experiments were performed on the data set presented in \cite{valentini_dataset}. The database is derived from the voice bank corpus \cite{vbn_corpus} from which recordings from $28$ speakers were chosen for the training set and $2$ for the test set. The recordings were added with $10$ different noise conditions ($2$ artificial and $8$ from the DEMAND database \cite{demand_database}) at signal-to-noise ratios (SNRs) of $0$, $5$, $10$ and $15$ dB. Thus the training set simulates $40$ different noisy scenarios and is comprised of a total of $11~572$ recordings. The test set was created using $5$ noise conditions (all from the DEMAND database, but different from training noise conditions) added at SNRs $2.5$, $7.5$, $12.5$ and $17.5$ dB. Altogether, the test set contains $824$ utterances. The database was downsampled from $48$ kHz to $16$ kHz for our experiments. 

\subsection{cGAN setup}
This work used $11$ convolutional layers each for $G_{\text{enc}}$ and $G_{\text{dec}}$ with filter-length $31$ and stride $=2$. Thus, after every layer, the temporal dimension of the features gets halved (Fig.  \ref{fig:generator}). We operated on signals that were sampled at $16$ kHz and considered approximately $1$ second of speech ($16~384$ samples) as input to the network. Thus, after $11$ convolutional layers in $G_{\text{enc}}$ the temporal dimension shrunk to $16~384/ 2^{11} = 8$. 

The number of feature-maps ($k_i$ in Fig. \ref{fig:generator}) used in the convolutional layers were: $16$, $32$, $32$, $64$, $64$, $128$, $128$, $256$, $256$, $512$ and $1024$. Thus the encoder output was of size $8\times 2014$ which was then concatenated with a latent vector of the same size. $G_{\text{dec}}$ followed the reverse procedure together with skip connections that doubled the temporal dimension after every layer resulting in a final output size that was identical to that of the input noisy signal.

As mentioned in Section \ref{sec:gan_framework}, $D$ used the same structure as $G_{\text{enc}}$, but with two input channels of $16~384$ samples each. The output of the convolutional layer (of size $8 \times 1024$) was fed to another convolutional layer of filter length 1 and stride 1, resulting in a representation of size $8 \times 1$. This was fed to a fully-connected layer for classification. 

The model was trained using the Adam optimizer \cite{adam_optimizer} (as opposed to RMSProp used in SEGAN \cite{santi_segan_is2017}) for $80$ epochs with a learning rate of $0.0002$ using a batch-size of $100$. The speech signals were windowed using sliding windows of length $16~384$ with $50$\% overlap. During testing, the enhanced signals were reconstructed by adding the generated signals with the same overlap and dividing the overlapping sections by $2$ to compensate for the $50$\% overlap. We also applied a pre-emphasis filter with $\alpha = 0.95$ to all input samples, except for the trainable pre-emphasis layer setting where the input to $G$ was not pre-emphasized.

\begin{table*}[ht!]
\caption{Comparison of the different GAN-based speech enhancement systems. Note that a higher value means a better performance for all the measures except CD and LLR. The best results obtained are highlighted in bold font. %The approaches are cumulative for the block with the $+$ sign, i.e., the additional components are added on top of the setting on the previous line. 
\label{tab:results}}
\centering
\begin{tabular}{l | c c c c c | c c c c c}
\toprule
\textbf{Setting} & \textbf{STOI} & \textbf{PESQ} & \textbf{CD} & \textbf{LLR} & \textbf{segSNR} &  \textbf{STOI} & \textbf{PESQ} & \textbf{CD} & \textbf{LLR} & \textbf{segSNR} \\
\midrule
Unprocessed & 0.921 & 1.97 & 4.41 & 0.46 & 8.77 & 0.921 & 1.97 & 4.41 & 0.46 & 8.77\\
LSTM-IRM \cite{dbaby_is2018} & 0.931 & 2.48 & \textbf{2.76} & 0.33 & 15.73 & 0.931 & 2.48 & \textbf{2.76} & \textbf{0.33} & 15.73 \\
\midrule
& \multicolumn{5}{c|}{\it With latent vector} & \multicolumn{5}{c}{\it Without latent vector}\\
\midrule
SEGAN \cite{santi_segan_is2017} & 0.928 & 2.16 & 3.35 & 0.48 & 15.69 & 0.925 & 2.18 & 3.39 & 0.44 & 15.43 \\ \midrule
IN & 0.933 & 2.49 & 3.11 & 0.43 & 16.66 & 0.934 & 2.50 & 3.11 & 0.44 & 16.45 \\
IN $+$ LabSmth & 0.938 & 2.53 & 3.04 & 0.43 & 17.01 & 0.938 & 2.54 & 3.20 & 0.34 & 16.68 \\
IN $+$ GT & \textbf{0.940} & 2.59 & 2.96 & 0.38 & 17.00 & 0.939 & \textbf{2.62} & 2.96 & \textbf{0.32} & \textbf{17.28} \\
IN $+$ PreEm & 0.939 & \textbf{2.64} & 3.06 & 0.37 & 17.13 & \textbf{0.941} & 2.57 & 3.20 & 0.35 & 16.42 \\
IN $+$ LabSmth $+$ GT $+$ PreEm  & 0.939 & 2.60 & 3.04 & 0.39 & \textbf{17.31} & \multicolumn{5}{c}{\emph{Unstable}}  \\
\bottomrule
\end{tabular}
  \end{table*}

Similar to \cite{santi_segan_is2017}, the $\lambda$ parameter that controls the L1 loss was set to $100$. The latent noise input $\mathbf{z}$ of size $8 \times 1024$ was drawn from a normal distribution $\mathcal{N}(\mathbf{0}, \mathbf{I})$. The whole project was developed in Keras \cite{keras} with Tensorflow \cite{tensorflow} back-end and is made available on github\footnote{The proposed cGAN-based speech enhancement framework is available at \tt{https://github.com/deepakbaby/isegan}}. 

\subsection{Evaluation metrics}
The speech enhancement performance was evaluated using the following measures: the short-term objective intelligibility (STOI) metric \cite{stoi}, perceptual evaluation of speech quality (PESQ) \cite{pesq} in terms of mean opinion score (MOS), segmental SNR (segSNR), cepstral distance (CD) and log-likelihood ratio (LLR). The CD, LLR and segSNR measures are expressed in dB and were obtained using the implementations provided with the REVERB challenge \cite{reverbchallenge}. Higher values of PESQ, STOI and segSNR, and lower values of CD and LLR indicate better performance.
\section{Results and Discussion} \label{sec:results}

The noise suppression performance obtained for the various speech enhancement systems in terms of various speech quality measures are provided in Table \ref{tab:results}. To compare cGAN with the conventional magnitude spectral enhancement approach, an LSTM-based speech enhancement system that estimates the ideal-ratio mask (IRM) for enhancing the Gammatone spectrogram of noisy speech is also included. The details of the model are provided in \cite{dbaby_is2018}. 

It can be seen that the LSTM-IRM model outperforms the SEGAN model, even though the former reused the noisy phase for reconstructing the time-domain signal. Using instance normalization (\emph{IN} in Table \ref{tab:results}) instead of VBN reduced the training time considerably and resulted in a better performance than using VBN. Using one-sided label smoothing (denoted as \emph{LabSmth}) together with instance normalization yielded a more stable GAN training and this approach outperformed the LSTM-IRM baseline model (except for CD and LLR).

Incorporating the trainable auditory filterbank (\emph{GT}) and pre-emphasis (\emph{PreEm}) layers yielded further improvements even without one-sided label smoothing. To our knowledge, this is the first time where such speech processing principles are incorporated in a cGAN-based speech enhancement system. The results show that this approach greatly benefits in stabilizing the model and yields improved state-of-the-art speech enhancement performance.

It can also be seen that using the latent vector $\mathbf{z}$ sometimes yield a better performance (\emph{with latent vector} vs. \emph{without latent vector}), suggesting that $G$ does not always learn to ignore the latent vector. Moreover, removing the latent vector made the last system unstable and the model failed to achieve any equilibrium. The effect of latent vectors in cGAN models requires further investigation, which is beyond the scope of this paper. In our experiments, the best performance was achieved using instance normalization with the trainable pre-emphasis layer and the latent vector. However, combining all the cGAN variants (\emph{InstNorm + LabSmooth + GTLayer + PreEmLayer}) yielded a slightly worse performance as it resulted in a different equilibrium state as compared to the other settings.
\section{Conclusions and Future Work} \label{sec:conclusions}
This paper investigated several approaches to improve the performance of end-to-end raw speech waveform enhancement using cGANs. First, we investigated using instance normalization instead of VBN in order to reduce the computational complexity during the model training. The investigated cGAN model that combined one-sided label smoothing with instance normalization was shown to outperform a VBN-based cGAN model and a LSTM-IRM-based speech enhancement system. To our knowledge, this is the first time a cGAN-based speech enhancement system is shown to outperform a popular IRM-based approach. This paper also showed that using trainable Gammatone-based auditory filtering and pre-emphasis layers also can stabilize the model as well as improve its performance. 

Since the proposed models are shown to outperform the popular IRM-based models, using the proposed cGAN-based models as a front-end for automatic speech recognition systems is a suggested future work. The project published in github offers more flexibility in terms of combining different auditory model-based initializations and other normalization approaches such as batch renormalization and group normalization. Investigating different combinations and applying them for dereverberation is also a promising research direction.

%\section*{Acknowledgements}
%This work was funded with support from the European Union's Horizon 2020 research and innovation programme under grant agreement No 678120 (RobSpear). We gratefully acknowledge the support of NVIDIA Corporation with the donation of the Titan Xp GPU used for this research.

\bibliographystyle{IEEEtran}
\bibliography{mybib}

\end{document}